\documentclass{aa}  

\usepackage{graphicx}
\usepackage{txfonts}
\usepackage{amsmath}
\usepackage{xcolor} 
\usepackage{ulem} 

\usepackage{natbib}

\begin{document}

   \title{SWEET-Cat 2.0: The Cat just got SWEETer}

   \subtitle{Higher quality spectra and precise parallaxes from GAIA eDR3}

   \author{S. G. Sousa\inst{1}                  
          \and V. Adibekyan\inst{1}             
          \and E. Delgado-Mena\inst{1}
          \and N. C. Santos\inst{1,}\inst{2}
          \and B. Rojas-Ayala\inst{3}           
          \and B. M. T. B. Soares\inst{1,}\inst{2}  
          \and H. Legoinha\inst{1,}\inst{2}          
          \and S. Ulmer-Moll\inst{4,}\inst{1}       
          \and J. D. Camacho\inst{1,}\inst{2}       
          \and S. C. C. Barros\inst{1}          
          \and O. D. S. Demangeon\inst{1,}\inst{2}
          \and S. Hoyer\inst{5}                 
          \and G. Israelian\inst{6}
          \and A. Mortier\inst{7,}\inst{8}            
          \and M. Tsantaki\inst{9}                
          \and M. Monteiro\inst{1}
          }

          \institute{Instituto de Astrof\'isica e Ci\^encias do Espa\c{c}o, Universidade do Porto, CAUP, Rua das Estrelas, 4150-762 Porto, Portugal
          \and Departamento de F\'isica e Astronomia, Faculdade de Ci\^encias, Universidade do Porto, Rua do Campo Alegre, 4169-007 Porto, Portugal
          \and {Instituto de Alta Investigaci\'on, Universidad de Tarapac\'a, Casilla 7D, Arica, Chile}
          \and Department of Astronomy, University of Geneva, 51 chemin Pegasi, 1290 Sauverny
          \and Aix Marseille Univ, CNRS, CNES, LAM, Marseille, France
          \and Instituto de Astrof\'isica de Canarias, 38200 La Laguna, Tenerife, Spain
          \and Astrophysics Group, Cavendish Laboratory, University of Cambridge, J.J. Thomson Avenue, Cambridge CB3 0HE, UK
          \and Kavli Institute for Cosmology, University of Cambridge, Madingley Road, Cambridge CB3 0HA, UK
          \and INAF -- Osservatorio Astrofisico di Arcetri, Largo Enrico Fermi 5, 50125 Firenze, Italy
          }

   \date{Received September 15, 1996; accepted March 16, 1997}

 
  \abstract
   {}
   {The catalog of Stars With ExoplanETs (SWEET-Cat) was originally introduced in 2013. Since then many more exoplanets have been confirmed, increasing significantly the number of host stars listed there. A crucial step toward a comprehensive understanding of these new worlds is the precise and homogeneous characterization of their host stars. Better spectroscopic stellar parameters along with new results from Gaia eDR3 provide updated and precise parameters for the discovered planets. A new version of the catalog, whose homogeneity in the derivation of the parameters is key to unraveling star--planet connections, is available to the community.
   } 
   {We made use of high-resolution spectra for planet-host stars, either observed by our team or collected through public archives. The spectroscopic stellar parameters were derived for the spectra following the same homogeneous process using ARES and MOOG (ARES+MOOG) as for the previous SWEET-Cat releases. We re-derived parameters for the stars in the catalog using better quality spectra and/or using the most recent versions of the codes. Moreover, the new SWEET-Cat table can now be more easily combined with the planet properties listed both at the Extrasolar Planets Encyclopedia and at the NASA exoplanet archive to perform statistical analyses of exoplanets. We also made use of the recent GAIA eDR3 parallaxes and respective photometry to derive consistent and accurate surface gravity values for the host stars.
   }
   {We increased the number of stars with homogeneous parameters by more than 40\% (from 645 to 928).  We reviewed and updated the metallicity distributions of stars hosting planets with different mass regimes comparing the low-mass planets (< 30M$_{\oplus}$) with the high-mass planets. The new data strengthen previous results showing the possible trend in the metallicity-period-mass diagram for low-mass planets.
   }
   {}


   \maketitle
%

\section{Introduction}

Studies of planetary systems have grown immensely in recent decades following the first detections of exoplanets orbiting solar-type stars \citep[e.g.,][]{Mayor-1995} and the thousands of exoplanets detected since. More interesting than the fantastic increase rate of discoveries is the surprising vast diversity of exoplanets present in the Galaxy, revealed by dedicated search surveys using the radial-velocity (RV) and transit techniques. 

These discoveries provide crucial constraints for the exhaustive understanding of planet formation and evolution, based initially on our knowledge of the Solar System  alone \citep[e.g.,][]{Ehrenreich-2020, Lillo-Box-2020, Armstrong-2020}. 

The reasonable assumption that the same source material forms the planets and host star(s) within a given planetary system has led researchers to focus on a quest for possible correlations between the characteristics of the host stars and the presence of their planets. One of the first observational results to become evident in this regard was that massive Jupiter-like planets are much more frequent around metal-rich stars \citep[e.g.,][]{Santos-2004b, Valenti-2005}. Later on, with further discoveries of lower mass planets, this planet--metallicity correlation was found to be nonexistent for small or low-mass planets \citep[e.g.,][]{Sousa-2008, Ghezzi-2010, Schlaufman-2011, Buchhave-2012, Buchhave-2015, Kutra-2020}. Several works have focused on other aspects, such as planetary system architectures \citep[][]{Adibekyan-2019, Dawson-2018},  and planet frequency, and its dependence on other stellar parameters like the stellar mass \citep[e.g.,][]{Mortier-2013b}. These are a small representation of the statistical studies performed focusing on the potential correlations between the star’s proprieties and its hosted planets. Detailed knowledge of the nature of individual stars is crucial for the thorough understanding of planetary systems currently being characterized by CHEOPS \citep[][]{Benz-2021} and ESPRESSO \citep[][]{Pepe-2021}, and for preparing the selection of targets for future missions such as ARIEL \citep[][]{Tinetti-2018, Brucalassi-2021}, and PLATO \citep[][]{Rauer-2014}. SWEET-Cat is a catalog with homogeneous spectroscopic parameters for planet hosts, with near-full completeness for the RV detected planets; it was introduced by \citet[][]{Santos-2013}. To increase the sample of stellar hosts with homogeneous spectroscopic parameters, our group has continuously worked on SWEET-Cat, releasing several updates to the community \citep[e.g.,][]{Sousa-2015a, Andreasen-2017, Sousa-2018}. 

This work presents a significant upgrade to SWEET-Cat\footnote{www.astro.up.pt/resources/sweet-cat which is now replaced by sweetcat.iastro.pt}. It consists of the addition of 283 new planet hosts, the inclusion of new data relevant for the characterization of the planet-host stars, the inclusion of the link to the NASA archive exoplanet database\footnote{http://exoplanetarchive.ipac.caltech.edu} \citep[][]{Akeson-2013}, in addition to the already existing link to the Extrasolar Planets Encyclopedia\footnote{http://exoplanet.eu/} \citep[][]{Schneider-2011}, and the rederivation of spectroscopic parameters from better quality or recent spectra using the latest updated codes developed by our group. The following sections describe all the work performed for the new version of SWEET-Cat. Section 2 describes the spectroscopic data compilation for the catalog. Section 3 presents how the spectroscopic parameters are derived, and compares the latest parameters with our older results and others in the literature. We also describe how we use the GAIA eDR3 to estimate accurate surface gravities for our planet-host stars. In Section 4 we present the new catalog contents and how we use the coordinates to link our catalog with the exoplanet.eu and the NASA archive. In Section 5 we review and update planet--star correlations discussed previously in the literature. Finally, in Section 6 we present the summary of our work.

\section{Spectroscopic data}

\subsection{Selection of host stars for SWEET-Cat}

We kept the same constraints of previous catalog versions to include the planet host stars in SWEET-Cat. First, we search for the confirmed planets in the Extrasolar Planets Encyclopedia and selected only the detected by Radial Velocity, Primary Transit, and Astrometry methods. Second, we cross-match their host stars with the stars in SWEET-Cat and include the missing hosts, while removing the stars that  no longer had a planet confirmed status. The same procedure is performed with the NASA exoplanet archive to add the confirmed host stars missing in the Extrasolar Planets Encyclopedia. We include the database source in the information of all host stars (see Section 4 for details).

\begin{figure}[t]
  \centering
  \includegraphics[width=8.5cm]{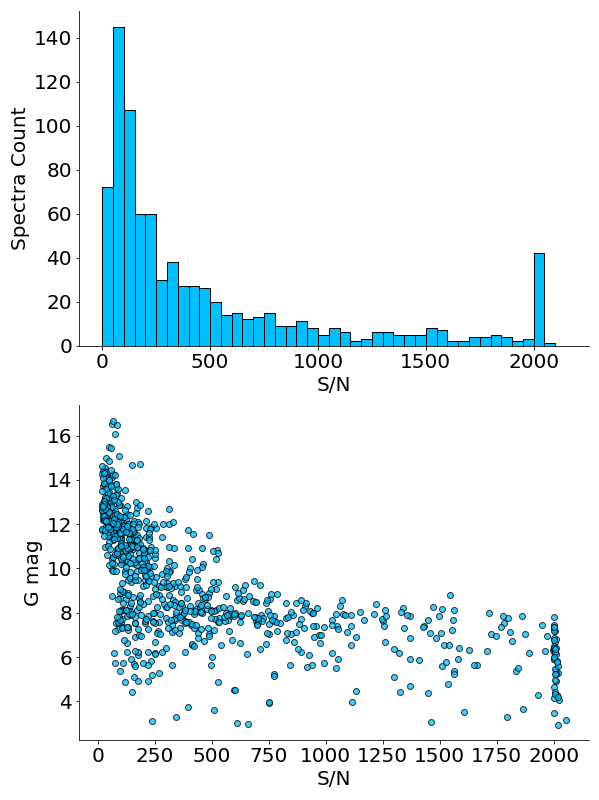} 
  \caption{S/N distribution and S/N vs. G magnitude diagram for SWEET-Cat spectral data (top and bottom panels, respectively).}
  \label{fig_sn}
\end{figure}

\subsection{Spectra from public archival data}

The ESO archive\footnote{http://archive.eso.org/wdb/wdb/adp/phase3\_main/form} is the main source of reduced spectral public data for SWEET-Cat. We used the \textit{astroquery.eso}\footnote{http://astroquery.readthedocs.io} submodule  to search for the required data in the ESO archive. We note that a small fraction of the spectra belonging to our UVES programs had private status at the moment of compilation (106.20ZM.001, 105.203J.001, 0103.C-0027, 0102.C-0226, 0101.C-0049).

The compilation of data prioritizes high-resolution and high signal-to-noise ratio (S/N) spectra for each host star. We selected the data from the FEROS, HARPS, and UVES spectrographs, all with R > 40000\footnote{ESPRESSO was not included here because it was not listed by astroquery.eso at the time of the compilation of the data.}.
In most cases we downloaded more than one spectrum per star to achieve a higher S/N by combining them. We avoided very low S/N spectra (S/N<15) due to low-quality individual exposures that could add relatively low signal compared with the noise. We also did  not select very high S/N spectra (e.g., S/N>550 for HARPS, S/N>1000 for UVES\footnote{The   S/N threshold for the case of saturation depends on the detectors of each spectrograph and/or configuration.}) either due to the likely saturation of the individual exposures. The total number of files downloaded varies per star to reach a maximum S/N of about 2000 for the final combined spectrum, and a minimum of 20 for a final  single-exposure  spectrum. We note that the cited S/N values are from photon counts and that the real S/N should depend on other noise sources (e.g., background subtraction, flat-fielding), although they should not differ much for the final combined spectrum. 
We obtain the final combined spectrum for each star for a given instrument and configuration by shifting all spectra to the same wavelength reference before the flux combination, using a cross-correlation function with the highest S/N individual spectrum taken as the mask.

We keep track of the search parameters used for the data in the archive (e.g., S/N cuts) and the list of the individual spectra used for the final combined spectrum. Moreover, we keep figures for each final combined spectrum along with the spectra used for its creation. 

Since this is a significant amount of data the selection of spectra and their combination is an automatic process. However, there is a human validation to certify the quality of the end products. The main problems identified during the validation include the following: 

i) incorrect identification of the star: some stars are very close in the sky. For example, stars in the Kepler field and close in binaries can be easily mismatched when searching the archival data. We identify mismatches by cross-checking the lists of spectra used for each star. if we find a file used for two or more stars, we discard that spectrum and redo the combination process with the updated list.

ii) data reduction process: some individual spectra in the figures show unexpected flux variations and/or strange behavior in the continuum (e.g., ripples or strong spikes). These likely indicate a problem with the reduction of the archival spectra. We discard these spectra and redo the combination process.  

iii) radial velocity correction: some spectra were not properly shifted in radial velocity before the spectral combination by looking at the figures. We identify the files and apply the cross-correlation function with different parameters to correct the issue. If that fails, we discard the files and redo the combination process.

We searched for spectra in other archives for the host stars of SWEET-Cat. In particular, archives where we could do the search and download automatically. We used custom-made python scripts to search for and download spectra from the SOPHIE\footnote{http://atlas.obs-hp.fr/sophie/}, ESPADONS\footnote{http://www.cadc-ccda.hia-iha.nrc-cnrc.gc.ca/en/
cfht/}, and HARPS-N\footnote{http://www.not.iac.es/archive/} spectrographs archival data. We combined these data in the same way as for the ESO spectral data.

We show the S/N distribution of the combined spectra and the dependence between the GAIA magnitude and the S/N in Fig \ref{fig_sn}. The peak at S/N $\sim$ 2000 arises because we stop the stacking of spectra once we reach this value.

\subsection{Legacy spectral data}

We keep the combined spectra from previous versions of SWEET-Cat for some host stars (75 spectra). They either are of better quality than the ones we combined automatically or are from a public archive of instruments not searched  in this version (e.g., FIES).
Also, we constrained the compilation of new data to host stars with temperatures above 4200K since we do not plan to use our methodology on cooler stars.

\subsection{Spectra format}

The spectra compilation followed the basic procedure  described in \citet[][]{Sousa-2018}. We store the combined spectra in fits files. In the fits header, in addition to the essential keywords, we added the search parameters, the GAIA DR2 id\footnote{At the time of the combination of spectra the GAIA DR2 was an identifier in Simbad (http://simbad.u-strasbg.fr/), while the GAIA eDR3 was not.}, the search box area, the date related to the data search, and the respective combinations. Moreover, we include the list with the reduced frames used for the combined spectra and their S/N information from the archives. We plan to provide such spectra in a future update of SWEET-Cat.

\section{Stellar parameters}

\begin{figure*}
  \centering
  \includegraphics[width=19cm]{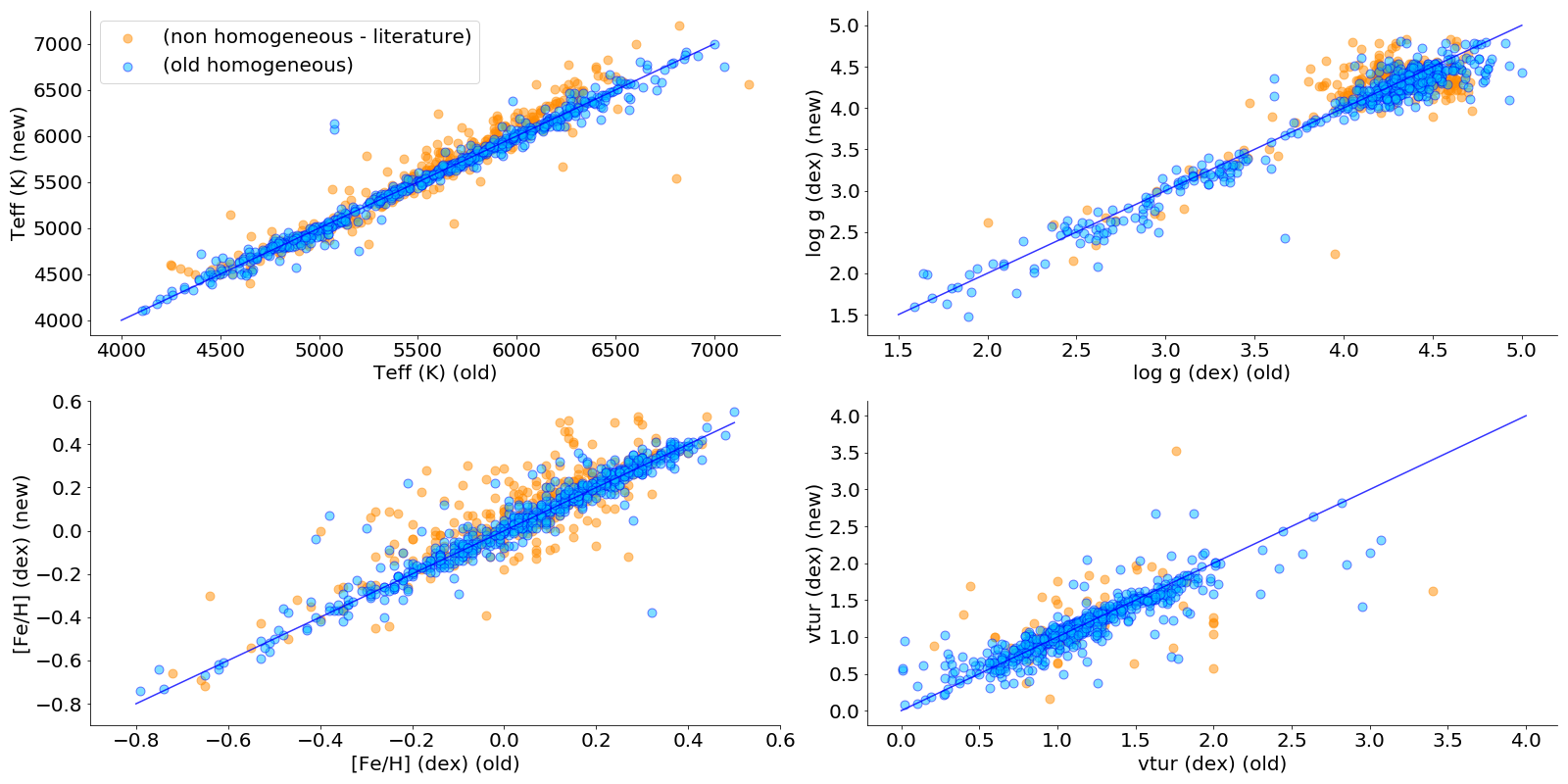}
  \caption{New SWEET-Cat homogeneous parameters ($T_{eff}$, $\log g$, $[Fe/H]$, and $v_{tur}$) compared to the literature values (orange filled circles) and the previous homogeneous values (blue filled circles) listed in SWEET-Cat.}
  \label{fig_param}
\end{figure*}

\subsection{Spectroscopic parameters}

We carried out the spectroscopic analysis with ARES+MOOG for the new spectroscopic data and the legacy data \citep[for details see][]{Sousa-2014}. We re-derived the homogeneous parameters already in SWEET-Cat for several host stars, given that we either collected higher quality data or decided to apply the latest version of the codes described previously. The spectral analysis relies on the excitation and ionization balance of iron abundance. The ARES code\footnote{The latest version of ARES can be found on github.com/sousasag/ARES} \citep[][]{Sousa-2007, Sousa-2015b} automatically measures the equivalent widths of the absorption lines. The MOOG code \citep[][]{Sneden-1973} is used for the element abundances assuming local thermodynamic equilibrium (LTE) and using a grid of Kurucz ATLAS9 plane-parallel model atmospheres \citep[][]{Kurucz-1993}. We applied this method in our previous spectroscopic studies of planet hosts \citep[e.g.,][]{Sousa-2008, Sousa-2011, Mortier-2013b, Sousa-2015a, Andreasen-2017, Sousa-2018}. We used the same line list introduced in \citet[][]{Sousa-2008}, except for the stars with effective temperature below 5200K where we used the line list provided in \citet[][]{Tsantaki-2013}.

The atomic data, in particular the oscillator strengths (log gf values), were re-calibrated to match the solar abundances as in our previous works, but in this case, we used the 2019 version of MOOG. The log gf values slightly differ, most likely due to numerical differences between the code compilations or versions.  

We collected combined spectra from two or more instruments or configurations for several host stars, and hence they have more than one set of converged parameters. The results from spectra from different instruments are generally consistent within the errors (see Fig. 3 in \citealt{Sousa-2008} for typical dispersion of parameters for the same stars with our methodology using spectra from different instruments). We include only one set of converged parameters per star in SWEET-Cat. The selection of such a set of parameters relies on the results of the spectroscopic analysis. The errors associated with the parameters are decisive since they are related to the quality of the spectra (S/N and resolution). If the errors are similar, we select the set of parameters from the higher resolution spectra. In Figure \ref{fig_param} we show the comparison between the parameters derived in this work and the values previously listed in SWEET-Cat or literature values.

In previous versions of SWEET-Cat 645 host stars had homogeneous parameters. The new and old parameters are consistent, regarding effective temperature and [Fe/H]; the mean differences are only -5±64 K and 0.01±0.06 dex, respectively. We derived homogeneous parameters for 278 new stars in SWEET-Cat. Consistency between our values and the literature values  (mainly gathered from the discovery paper) is not as good; for effective temperature and [Fe/H] the mean differences are 62±181 K and 0.05±0.14 dex, respectively. There are significant differences between the parameters for some host stars, even when they were already in SWEET-Cat. The use of higher quality spectra for the new parameters can explain the majority of these discrepancies. We exemplify
other reasons with the following two stars. For HAT-P-23, we had parameters from a lower S/N spectrum using a synthesis method, but since we recovered a  higher quality spectrum and the star has a reasonable low vsini, we now used the standard ARES+MOOG methodology. For HIP75458, there was an undetected typo for the listed effective temperature in SWEET-Cat. Finally, these comparisons identified mismatches in the spectrum source. For example, we got a completely different set of parameters for HD98219 using the new UVES spectra compiled from the ESO archive. We then realized that the downloaded spectra belonged to TYC 6649-793-1, and therefore we kept the values from the FEROS legacy spectrum used in previous works for HD98219.

\subsection{Use of GAIA eDR3}

We matched the stars listed in SWEET-Cat with the GAIA ID in DR2 and eDR3 using their coordinates and the VizieR catalogs \citep[][]{GAIA-2016, GAIA-2021}. We checked the magnitudes and astrometry data to confirm that we selected the correct star. While most stars in SWEET-Cat have the same ID as for GAIA DR2, there are GAIA IDs that appear to change  between the data releases. The correct GAIA ID identification was not straightforward in some cases because of the presence of other stars with similar magnitudes in the searched field of view. The GAIA IDs make it easier for SWEET-Cat users to combine the spectroscopic data with the GAIA astrometry. For each target with a GAIA ID, we proceed to the extraction of its parallax from eDR3. Parallax was one of the columns present in the SWEET-Cat table, but not necessarily updated for the whole sample. GAIA parallaxes should be taken with caution for very bright stars as they may have additional systematic errors due to calibration issues \citep[][]{Drimmel-2019}. We included in the SWEET-Cat table the parallax listed in Simbad, if available, with the respective flag for the very bright host stars with no GAIA data. We have also included in  SWEET-Cat the  relevant columns for stellar characterization from GAIA photometry, such as the G magnitude (Gmag), the red and the blue passband magnitudes (RPmag and BPmag), and the mean flux for the G band ($FG$). We included a photometric activity index based on the GAIA photometry variability

\begin{equation}
 G\_flux\_std\_n = \frac{e\_FG}{FG}*\sqrt{oGmag},
\end{equation}which is the relative error on the G mean flux, where $oGmag$ is the number of observations (CCD transits) that contributed to the G mean flux and mean flux error ($e_FG$).

\subsection{Precise surface gravity from GAIA parallax}

We recall that the spectroscopic stellar surface gravities derived by our method are the least constrained parameters in our analysis \citep[e.g.,][]{Tsantaki-2019}. The reason behind this was addressed in our past works and is related to the small number of ionized iron lines available in the optical spectrum that we can measure. We  introduced some corrections to improve the surface gravity estimations \citep[e.g.,][]{Mortier-2014}. An alternative is to estimate the trigonometric surface gravity, as we did in the past \citep[e.g.,][]{Santos-2004b, Sousa-2008, Tsantaki-2013, DelgadoMena-2017}, using the following expression derived from the luminosity-radius-temperature relation for stars:

\begin{figure}[ht]
  \centering
  \includegraphics[width=8.5cm]{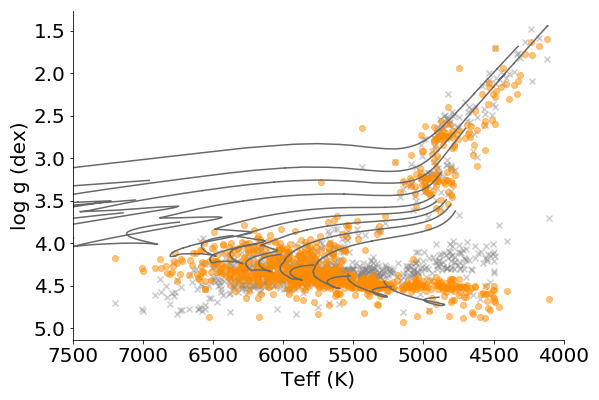} 
  \caption{Surface gravity vs. effective temperature as a proxy of the Hertzprung--Russell diagram. The black lines represent CESAM evolution model tracks for stellar masses ranging from 0.8 to 2.4M$_\odot$. The trigonometric surface gravities computed using the GAIA eDR3 parallaxes (orange circles) are in better agreement with the model tracks than the spectroscopic values (gray crosses) for the dwarf stars.}
  \label{fig_hr}
\end{figure}

\begin{equation}
 \log \frac{g}{g_{\odot}} = \log \frac{M}{M_{\odot}} + 4\log \frac{T_{eff}}{T_{eff\odot}} - \log \frac{L}{L_{\odot}}
,\end{equation}where $g$ is the stellar surface gravity, $M$ is the mass, $T_{eff}$ is the effective temperature, and $L$ is the luminosity. We derived the stellar luminosities directly from the precise data in GAIA eDR3, following  equation 8.6 from the GAIA documentation\footnote{https://gea.esac.esa.int/archive/documentation/GDR2/, in particular chapter 8.3.3 authored by Orlagh Creevey and Christophe Ordenovic.}

\begin{equation}
 -2.5 \log \frac{L}{L_{\odot}} = M_G + BC_G(T_{eff}) - M_{bol\odot}
,\end{equation}where $M_G$ is the absolute GAIA magnitude, $T_{eff}$ is the effective temperature listed in SWEET-Cat, $M_{bol\odot} = 4.74$ (as defined by the IAU resolution 2015 B2), and $BC_G(T_{eff})$ is a temperature-dependent bolometric correction provided in the GAIA documentation (equation 8.9 and Table 8.3, \citealt[][]{Andrae-2018}). Instead of a simple inversion of the GAIA parallax to provide a distance, we preferred to include the geometric distance and an error directly taken from the maximum of the asymmetric uncertainty measures (16th and 84th percentiles) reported in \citet[][]{Bailer-Jones-2021}.

\subsection{Precise surface gravity from GAIA parallax}

To estimate the trigonometric surface gravity, we need the stellar mass. As in previous works, we used the stellar mass calibration in \citet[][]{Torres-2010}, and for estimates between 0.7 and 1.3$M_\odot$ we used the correction in \citet[][]{Santos-2013}. However, this \citet[][]{Torres-2010} calibration requires spectroscopic parameters of the star as input, including the surface gravity. Therefore, we performed an iterative process to converge simultaneously for the best estimates of the stellar mass and the trigonometric surface gravity. The procedure consists of the following steps:

\begin{figure}[ht]
  \centering
  \includegraphics[width=8.5cm]{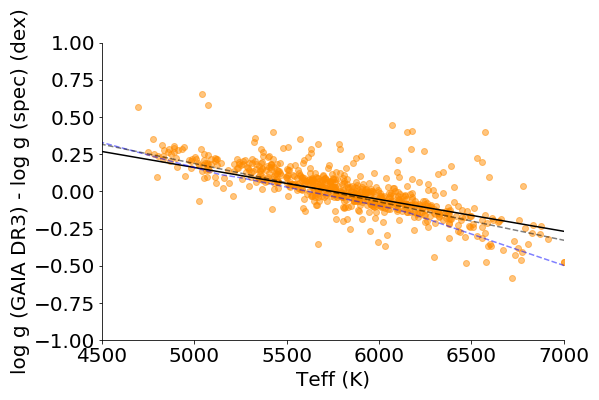}
  \caption{Difference between the trigonometric and the spectroscopic surface gravities as a function of effective temperature for dwarf stars (log g > 4.0 dex) in SWEET-Cat. The black lines show the linear fit represented in Equation \ref{eq4} , while the gray and blue dashed lines are fits presented in Fig. 5 of \citet[][]{Tsantaki-2013} and Fig. 2 of \citet[][]{DelgadoMena-2017}, respectively.}
  \label{fig_logg}
\end{figure}

\begin{enumerate}
\item The spectroscopic surface gravity is used as an initial value;
\item The surface gravity, together with the $T_{eff}$ and the $[Fe/H]$, is then used to compute the stellar mass from the \citet[][]{Torres-2010} calibration;
\item The stellar mass is used to re-derive a new trigonometric surface gravity;
\item If the new surface gravity is within 0.01 dex of the previous value, we stop this process. Otherwise, we go back to step 1 with the new surface gravity value for the next iteration.
\end{enumerate}

We applied the same methodology for the trigonometric surface gravity and the final stellar mass errors as in previous works \citep[see][]{Santos-2013, Sousa-2018}. We used a Monte Carlo approach taking 10\ 000 random values from the Gaussian distributions of the input parameters considering their uncertainties. We estimated the errors from the standard deviation of the stellar mass and the trigonometric surface gravity distributions in the final iteration. A potential flaw in this iterative procedure can be the lack of interaction with the other spectroscopic parameters (effective temperature and [Fe/H]) since they are also inputs for the mass estimation. However, from our experience with ARES+MOOG, we know that there is low interdependence of these parameters and in particular with the surface gravity (this is consistent with other equivalent-width methods using MOOG as discussed in \citealt[][]{Torres-2012} and \citealt[][]{Mortier-2014}). Using a subsample of the planet-hosts in SWEET-Cat, we show in Figure \ref{logg_fixed_deps} the small differences that we find for the effective temperature and metallicity when constraining the surface gravity with the trigonometric value for the spectroscopic analysis. Hence, the spectroscopic effective temperature and metallicity are well constrained and do not change significantly with the potential change of the surface gravity. Figure \ref{fig_hr} shows the SWEET-Cat stars in the log g -- Teff diagram, along with solar metallicity CESAM stellar model tracks for main-sequence to post-main-sequence stars with masses ranging from 0.8 to 2.4 $M_\odot$ \citep[][]{Marques-2008}. The trigonometric surface gravities agree better with the theoretical tracks than with the spectroscopically derived values, especially for cool dwarf stars.

\subsection{Correction for the spectroscopic surface gravity}

We can correct our spectroscopic surface gravity if we take the trigonometric surface gravity as the true value of the stellar surface gravity. The difference between the trigonometric and spectroscopic surface gravity values depends on the effective temperature of the stars for dwarf stars. We show this difference as a function of effective temperature for dwarf stars (logg > 4.0) in Figure \ref{fig_logg}. Considering the trend seen in Figure \ref{fig_logg}, we use a weighted linear fit to get a corrected spectroscopic surface gravity $\log g_{\textrm{c}}$ from the spectroscopic surface gravity and effective temperature: 

\begin{equation}
 \log g_{\textrm{c}} = \log g_{\textrm{s}} + ( 1.23 \pm 0.06) - (2.15 \pm 0.11)\times10^{-4} T_{\textrm{eff}}
\label{eq4}
.\end{equation}This linear relation, although not applicable to all the catalog stars, is not that far from those reported in \citet{Tsantaki-2013}, \citet{Mortier-2014}, and \citet{DelgadoMena-2017}.

Although we trust that the derived spectroscopic surface gravity does not affect the rest of the spectroscopic parameters by ARES+MOOG, the corrected loggs values can aid in the identification of the evolutionary status of the star by providing constraints for its age. Accurate surface gravities may be also relevant for the chemical abundances that rely on few spectral lines with a strong dependence on log g.

\section{catalog contents}

\begin{table}[t]
\caption[]{Updated statistics for SWEET-Cat.}
  \begin{center}

    \begin{tabular}{ll}
    \hline
    \hline
    Number     &   Description  \\
    \hline

3236 & stars in SWEET-Cat  \\
928  & planet hosts with homogeneous parameters \\
283  & new stars with homogeneous parameters \\
233  & bright FGK stars without \\
     & \ \ homogeneous parameters (G $<$ 12) \\
    \hline
    \end{tabular}
  \end{center}
\label{tab_stats}
\end{table}

\subsection{SWEET-Cat planet-host statistics}

Table \ref{tab_stats} presents the updated numbers of host stars in SWEET-Cat. The main reason for a relatively small number of stars with spectra (and therefore homogeneous parameters\footnote{The column ``SWFlag'' lists the stars for which we have derived spectroscopic parameters in SWEET-Cat. This flag was called ``source'' in the old table.}) is that a considerable number of the planet-hosts come from the detections provided by the Kepler mission \citep[][]{Borucki-2010}. These are faint stars and are usually not followed up by RV surveys since it is observationally costly to collect high-quality high-resolution spectra for them. For bright stars the spectroscopic stellar parameters are fundamental to get precise bulk stellar parameters. The stellar mass and radius are crucial to obtain reliable planetary-mass measurements from RV surveys and reliable planetary radius measurements from precise transit observations \citep[e.g.,][]{Hoyer-2021, Leleu-2021}. The percentage of stars with spectra and homogeneous parameters increases for bright stars listed in SWEET-Cat. With the work presented in this paper, considering the addition of 283 new stars with homogeneous parameters, we have $\sim$88\% and $\sim$76\% completeness for stars with G<9 and G<12, respectively. Completeness represents the number of host stars with homogeneous spectroscopic parameters relative to the total number of stars listed in SWEET-Cat. We significantly increase the number of homogeneous parameters for planet host stars with this update. However, the continuous increase of new host stars in the catalog keeps the percentage for bright stars almost the same. In Table \ref{tab_stats} we can also see that there is still a subset of bright planet-host stars (233 stars, about $\sim$25\% of the bright FGK stars) for which we did not find, at the time of the compilation of this data, high-quality public data at the time of the search for this work. These will be the focus of the next SWEET-Cat update.

\subsection{Online table}

We add new parameters in the updated version of SWEET-Cat\footnote{URL: sweetcat.iastro.pt}, and their descriptions are in Table \ref{tab_contents}. In the web page version of the catalog only a fraction of these columns are present for visualization reasons, but all columns are in the table available for download at the website. On the new website the community can find additional content: a Python notebook example on how to use the new SWEET-Cat table and a gallery with updated figures of the population of the planet-host stars and the correlations found between the planets and hosts stars.

\subsection{Link to exoplanet catalogs}

The match of SWEET-Cat host star parameters with the properties of their exoplanets in online databases is quite useful for the statistical exploration of the data. However, given that we only control the SWEET-Cat table, the continuous match with external databases is not always straightforward. At first SWEET-Cat kept entries for the confirmed planet-hosts in exoplanet.eu exclusively, and we did the matching with the names of the planets that inherit it from the host star. This approach turned out to be problematic:  some planets  names  either changed regularly (Keppler objects of interest to Kepler denominations is a typical example) or were not consistent within the planetary system (e.g., K2-50 with K2-50b and EPIC201833600 c listed in exoplanet.eu). Therefore, we now   rely on the host coordinates obtained from the Simbad database, and we compare them against the coordinates listed in exoplanets.eu and the NASA exoplanet archive. The updated SWEET-Cat table has the planet coordinates listed in both the exoplanet.eu and the NASA archive exoplanets database to facilitate any matching. By adding the planet coordinates in both databases, we can match the objects even if the coordinates are incorrect (e.g., due to typos) in the planet databases. We provide a python script example on the website to show the capabilities to explore SWEET-Cat and the properties of the exoplanets statistically.

\section{Exoplanets and their host stars: Review of star--planet relations}

\subsection{Revisiting the planet frequency--stellar metallicity correlation}

\begin{figure*}
  \centering
  \includegraphics[width=18cm]{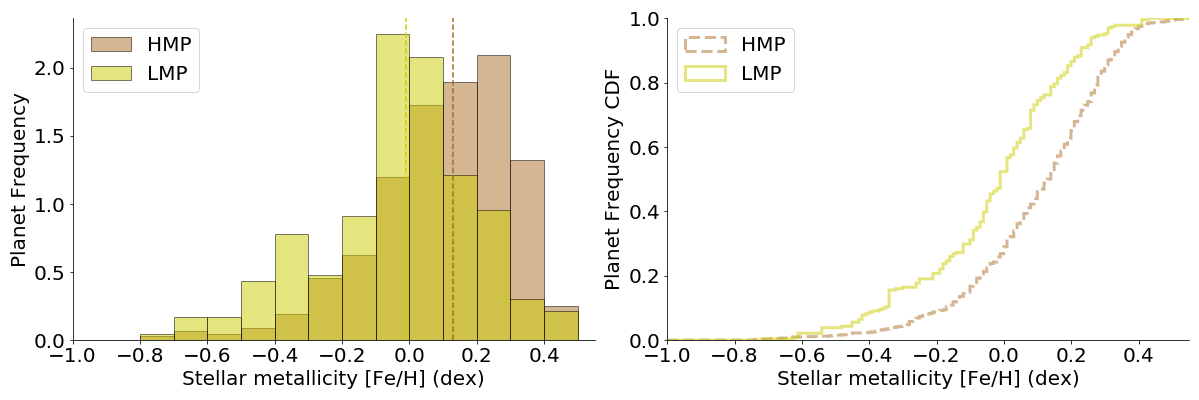}
  \includegraphics[width=18cm]{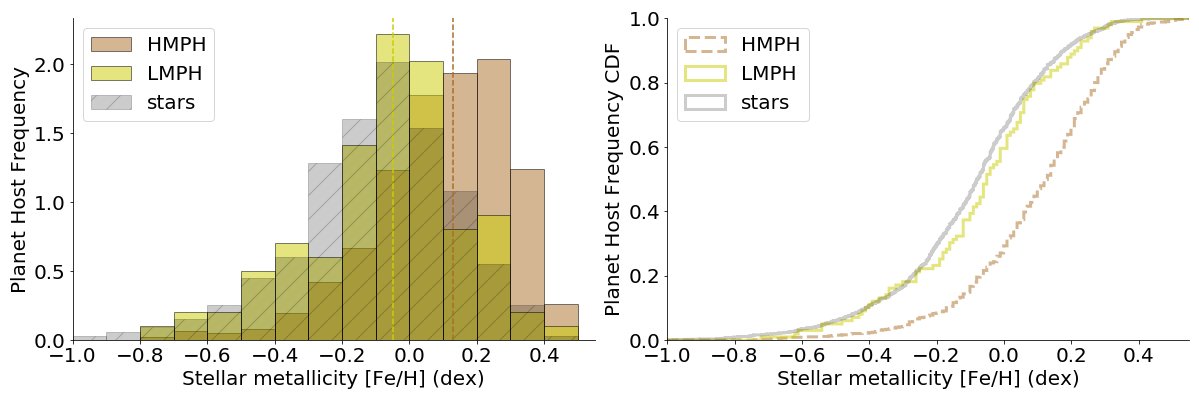}
  \caption{Metallicity distributions for the high-mass and low-mass planet samples (HMP and LMP, respectively). The top and bottom left panels show the metallicity distribution per planet and host star, respectively. The dashed vertical lines indicate the average of each distribution. The right panel shows their CDFs for a better comparison.}
  \label{planet_metallicity_hist}
\end{figure*}

The strong correlation between the presence of planets and the stellar metallicity observed for the host stars is the first observational constraint for planet formation theories. It became evident soon after the first discoveries of exoplanets \citep[e.g.,][]{Gonzalez-2000, Santos-2001, Santos-2004b, Fischer_Valenti-2005}. However, these first detections suffered a strong observational bias: the most massive and hotter planets are easier to be discovered. The improvement in the precision of the detection methods allowed the discovery of lighter and cooler planets, increasing the planetary system sample to study the planet-metallicity correlation. In particular, the planet--metallicity correlation seemed not to be present even for the very first detected low-mass exoplanets, or to be at least different from the positive correlation observed for giant planets \citep[e.g.,][]{Udry-2006, Sousa-2008, Ghezzi-2010, Sousa-2011, Buchhave-2012}. \citet[][]{Wang-2015} suggested the possibility of a universal planet--metallicity correlation for all planets, but it could be related to a higher planet frequency and a lower detectability of low-mass planets \citep[see also ][]{Zhu-2016}. Today the planet--metallicity correlations seem to be well established in the exoplanet community, but the detection of additional low-mass planets with precise masses can aid in the understanding of these correlations in different regimes. SWEET-Cat is a perfect test bench for these studies due to its homogeneous nature \citep[e.g.,][]{Adibekyan-2019} using the previous version of SWEET-Cat.

We present the metallicity distributions for two different populations   classified according to their masses: low-mass planets (LMP, <30 M$_\oplus$: 231 planets) and low-mass planet hosts (LMPH: 99 planet hosts), and high-mass planets (HMP, >30 M$_\oplus$: 889 planets) and high-mass planet hosts (HMPH: 771 planet hosts) with homogeneous parameters from SWEET-Cat in Figure \ref{planet_metallicity_hist}. We kept  30 M$_\oplus$ as the split reference value of the samples of exoplanets to match the location of the gap in the planet mass distribution in \citet{Mayor-2011} and to be consistent with our previous works \citep[e.g.,][]{Sousa-2011b, Sousa-2018}.  Figure \ref{planet_metallicity_hist} is an updated version of Figure 4 in \citet[][]{Sousa-2018}. In the top panel we plot the metallicity distribution for all planets, and therefore the stellar metallicity is repeated more than once for the multi-planetary systems. In the bottom panel we plot the metallicity distribution for the most massive planet in each system only. For comparison, we include the metallicity distribution of solar neighborhood stars from both the HARPS and CORALIE volume limited-based samples (discussed in \citealt[][]{Sousa-2011})  (see also \citealt[][]{Udry-2000, Santos-2004b, Sousa-2008}). The positive metallicity correlation for massive planets continues to hold, even with the inclusion of new planetary systems in the distribution. We used the Kolmogorov--Smirnov (K-S) test to quantify the similarities of the three populations studied here.  Table \ref{tab_stats_ks} contains the statistics for the comparisons. The K-S test results confirm that it is unlikely that the LMPH and HMPH populations belong to the same stellar sample and cannot rule out that the LMPH and solar neighborhood populations may belong to the same population. However, in the  bottom right panel of Figure \ref{planet_metallicity_hist} there is a slight shift of the LMPH CDF toward metal-rich values compared with the CDF of the neighborhood stars. We did not observe this shift in \citet[][Figure 4]{Sousa-2018}, where the CDFs were indistinguishable from each other. This behavior may be linked to the slight planet--metallicity correlation for low-mass stars discussed in \citet[][]{Sousa-2019} and later in this paper in section 5.3.

\begin{figure*}
  \centering
  \includegraphics[width=18cm]{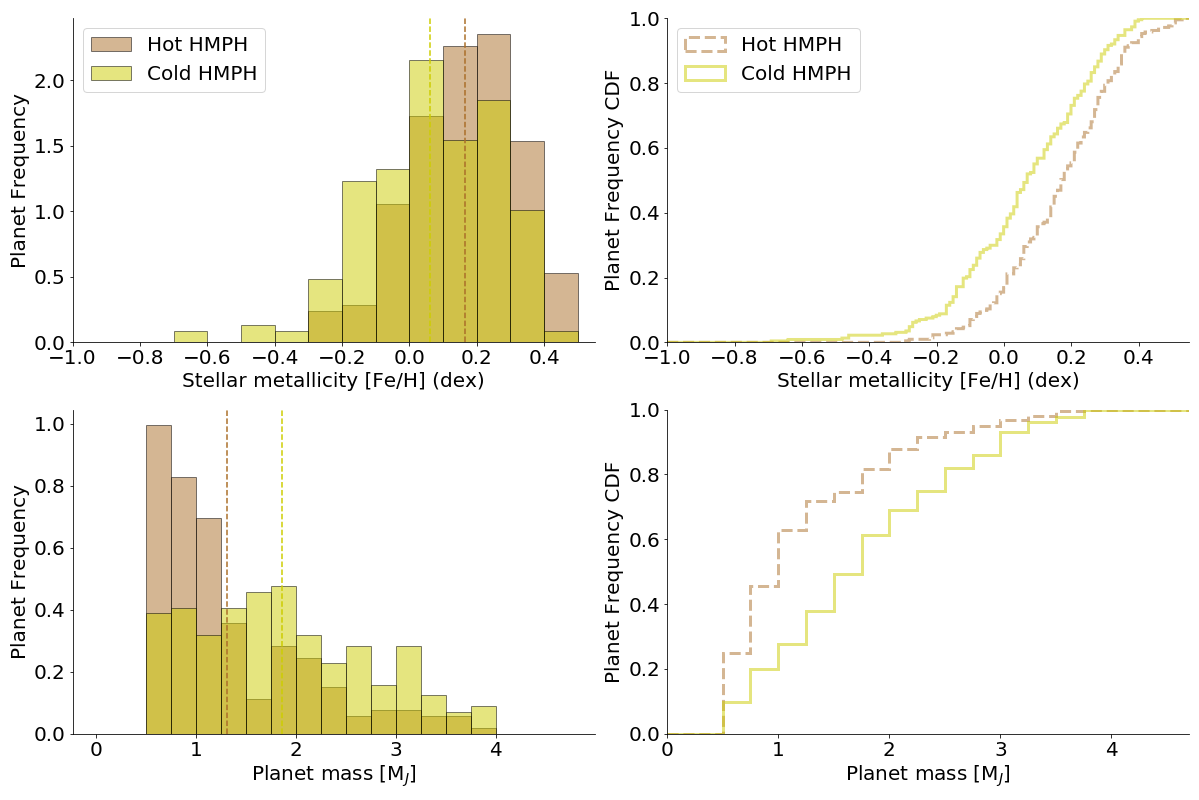}
  \caption{Metallicity and planet mass distributions for the hot- and cold-Jupiter samples (top and bottom left panels, respectively). The dashed vertical lines indicate the average of each distribution. The right panel shows their CDFs for a better comparison.}
  \label{hot_cold_hist}
\end{figure*}



\begin{table}
\caption[]{Kolmogorov--Smirnov test comparing different distributions.}
\small
  \begin{center}

    \begin{tabular}{lccc}
    \hline
    Samples        &      K-S        &   K-S   & mean    \\
                   &   statistic     & p-value & difference  \\
    \hline
    \\
    $[Fe/H]$ distributions: \\
    \\
    HMP vs. LMP (1)     & 0.29 & 2.83e-14      & 0.134\\
    HMPH vs. LMPH  (2)  & 0.35 & 4.03e-10      & 0.168\\
    LMPH vs. stars (2)  & 0.09 & 3.55e-01      & 0.034\\
    Hot HMPH vs. Cold HMPH (1) & 0.23 & 1.76e-05 & 0.104\\ 
    \hline
    \\
    M$_J$ distributions: \\
    \\
    Hot HMPH vs. Cold HMPH (2) & 0.36 & 1.99e-13 & -0.54\\
    \hline
    \\
    M$_\odot$ distributions: \\
    \\
    HMPH vs. LMPH (2)          & 0.40 & 1.95e-12 & 0.176\\
    Hot HMPH vs. Cold HMPH (2) & 0.29 & 6.02e-07 & 0.098\\
    \hline
    \end{tabular}
  \end{center}
\label{tab_stats_ks}
\tablefoot{(1) - Using planet counts; (2) - Using star counts. (see Fig. \ref{planet_metallicity_hist}); ``stars'' corresponds to the solar neighborhood sample.} 
\end{table}

\subsection{Hot vs. cold Jupiters}

\begin{figure}[t]
  \centering
  \includegraphics[width=9cm]{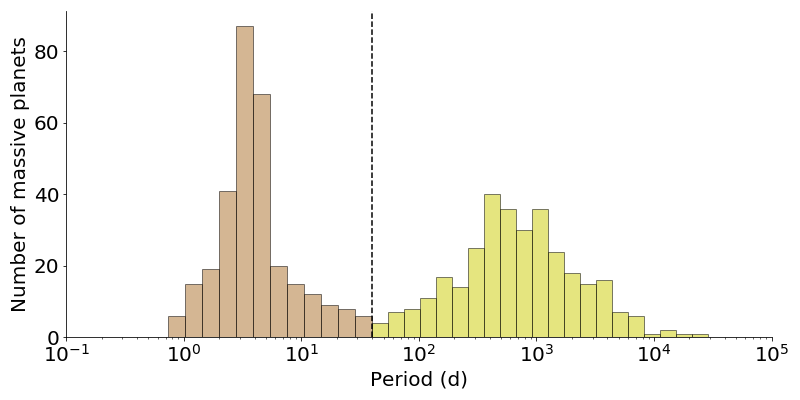} 
  \caption{Period distribution of massive planets with masses between 0.25 M$_J$ and 4 M$_J$. The black dashed line indicates the period (40 days) used to separate the two populations in this work.}
  \label{fig_hm_period_dist}
\end{figure}

\begin{figure*}
  \centering
  \includegraphics[width=18cm]{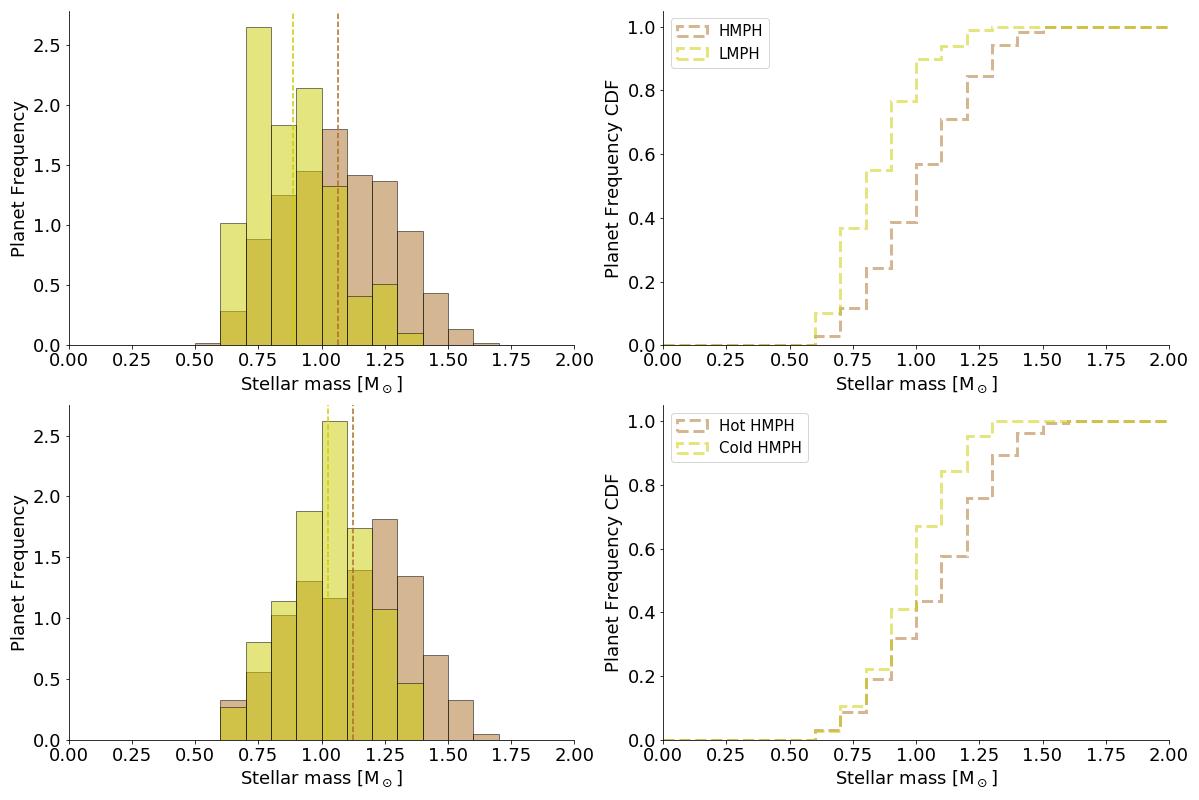}
  \caption{Stellar mass distributions for the high- and low-mass planet samples (HMPH and LMPH, top panel) and the hot and cold massive planets (Hot HMPH, and Cold HMPH, lower panel). The distributions are limited to host stars with log g > 4.0 dex. The dashed vertical lines indicate the average of each distribution. The right panel shows their CDFs for comparison.
}
  \label{stellar_mass_hist}
\end{figure*}

The increased precision of the detection methods allows the discovery of lower mass planets, but the course of time works in our favor to constrain the frequency of Jupiter-like planets found at long periods. We  focus now on the massive planets,  with masses above one-quarter of a Jupiter mass (>0.25 M$_J$, equivalent to $\sim$80 M$_\oplus$), and split them between hot and cold HMPs using a period cut at 40 days. We  also discarded planets with masses above 4 M$_J$, as in \citet[][]{Santos-2017}, to avoid overlap with possible brown dwarfs (most values are the minimum mass msini). The 40-day split value of the sample of HMPs follows from the double-peaked period distribution in Figure \ref{fig_hm_period_dist}. We show the metallicity distribution of the hot and cold populations of massive planets in Figure \ref{hot_cold_hist}. Although both groups are metal-rich compared with that of field dwarfs, contributing both similar to the metal-rich nature of stars hosting HMPs, there are some differences in the profile of their distribution. The K-S test to quantify the similarities is presented in Table \ref{tab_stats_ks} and indicates a low probability that both populations belong to the same sample of stars. In particular, cold HMPs are more frequent around metal-poor stars than hot HMPs. Known hot-HMP hosts exhibit metallicities higher than -0.4 dex, while the metallicity distribution of cold-HMP hosts reaches lower values ([Fe/H]$\sim$-0.7 dex). These results agree with the period-metallicity correlation \citep[see][]{Adibekyan-2013, Beauge-2013, Mulders-2016, Wilson-2018, VanDerMarel-2021}. The mean metallicity difference between hot and cold HMPs is 0.09 dex, exhibiting the same trend as the 0.16 dex difference observed in \citet[][]{Mulders-2016} and \citet[][]{Wilson-2018}. The value discrepancy may be related to the use of different samples of stars, and/or the split selection of the HMP. In \citet[][]{Mulders-2016} and \citet[][]{Wilson-2018} the planet sample contains both the most and the less massive planets, while we isolate them in a specific range of masses (0.25 - 4 M$_J$). These works propose possible explanations to explain this metallicity enrichment for short-period planet hosts: dust sublimation, which can correlate with metallicity \citep[e.g.,][]{Muzerolle-2003}; the planet migration mechanism and its dependence on metallicity, where planets around higher metallicity stars are ``trapped'' closer to the host stars while migrating inward \citep[e.g.,][]{Plavchan-2013}; rocky planet ingestion in the host star, forced by the migration of the short-period giant planets \citep[e.g.,][]{Mack-2014}; and migration by planet-planet interactions \citep[][]{Dawson-2013}. \citet[][]{Wilson-2018} address all of these and other possibilities, but it is still unclear which is responsible for this observational result. We also cannot rule out that several of these mechanisms can work simultaneously at some stages of the evolution of the systems. In the bottom panel of Figure \ref{hot_cold_hist}, we show the distribution of planetary mass for both cool and hot HMPs. Hot Jupiters are a distinct population from the cool Jupiters, exhibiting a right-skewed mass distribution with a median close to the Jupiter mass. The cold HMPs show a flat distribution between 0.25 - 2 M$_J$, skewing toward the right for masses higher than 2 M$_J$. However, we must keep in mind that it is more difficult to detect planets at long periods, and therefore we might be missing the detection of planets with masses lower than  or similar to that of Jupiter at long periods. As suggested by \citet[][]{Santos-2017}, different formation mechanisms acting for these populations can explain the discrepancies in mass distribution. For example, many of these more massive bodies may have formed as low-mass stars do \citep[e.g.,][]{Luhman-2012}. \citet[][]{Adibekyan-2019} suggests that planets of the same mass can form through different channels depending on the mass of the star--disk (i.e., environmental conditions can affect planet formation). Another possible explanation for the mass distribution difference may be related to an in situ formation hypothesis that may preferentially produce lower mass hot Jupiters. In this scenario, a massive planet close to the star will have fewer local solids and less gas to accrete, setting by default an upper mass limit related to the local disk structure \citep[e.g.,][]{Batygin-2016}. Migration can also explain this discrepancy if we consider that a more massive body can open deep gaps in the disk, decreasing the migration efficiency  \citep[][]{Masset-2003}. In addition, the evaporation of gas giants close to the star may remove enough material to decrease their initial mass value. For a review of hot Jupiters, see  \citet[][]{Dawson-2018} and references therein. 
 
\subsection{Stellar mass and planetary mass}

The relation between the occurrence rate of planets and stellar mass is a statistical result from exoplanetary surveys that enriches our understanding of planet formation \citep[e.g.,][]{Fulton-2018, Lozovsky-2021, Neil-2018, Yang-2020}. Because SWEET-Cat is a catalog of host stars only, it is impossible to estimate the occurrence rates of planets and their dependence on stellar parameters. However, SWEET-Cat data can be used to check if the mass distributions of stars hosting different planet types agree with the results related to occurrence rates. We focus these analyses on the hosts with trigonometric surface gravities above 4.00 dex (dwarf stars) since we rely on the \citet[][]{Torres-2010} calibration. As we do in Section 5.1., we split the sample into low-mass planet hosts LMPHs and high-mass planet hosts HMPHs, and we show their distributions in the top panels of Figure \ref{stellar_mass_hist}. The LMPHs present, on average, lower masses than the HMPHs. The LMPH and HMPH distributions have averages of 0.89 $\pm$ 0.16 M$_\odot$ and 1.06 $\pm$ 0.21 M$_\odot$, respectively. The K-S  test indicates a very low probability (1.95e-12) that the samples are drawn from the same distribution. The mass distributions of LMPHs and HMPHs in SWEET-Cat agree with surveys results that indicate that the occurrence rate of sub-Neptunes increases as stellar mass decreases \citep[e.g.,][with spectral type as a proxy for stellar mass]{Mulders-2015}. However, low-mass or small planets are harder to detect around massive dwarf stars, especially at large distances, and we may be missing them. This observational bias could therefore play a role in the shape of the sub-Neptune distribution.  
If we now focus only on the HMPHs and classify them by the type of planet they have (cold or hot), we see a marginal difference in their respective stellar mass distributions (bottom panels of Figure \ref{stellar_mass_hist}). The stars with cold and hot HMPs have a mean mass of 1.02 $\pm$ 0.16 M$_\odot$ and 1.12 $\pm$ 0.22 M$_\odot$, respectively. The K-S test assigns a low probability (6.02e-7) that these samples are drawn from the same distribution. Interestingly, the mass distribution of the hosts of hot HMPs is slightly double-peaked, and up to $\sim$ 0.9 M$_\odot$, both distributions are similar. Once again, observation biases may play a role in what we see since massive long-period planets are easier to detect around low-mass host stars.


\subsection{Metallicity-period-mass diagram for low-mass planets}

In \citet[][]{Sousa-2019}, we suggested a hint for a correlation in the metallicity-period-mass diagram for low-mass exoplanets (<30 M$_\oplus$), where low-mass exoplanets have increased masses for longer periods at higher metallicities. However, the observational bias from RV surveys where the massive planets are easier to discover at long periods could explain the correlation. The sample studied in \citet[][]{Sousa-2019} contained 84 exoplanets with homogeneous metallicities and mass determination with a relative precision within 20\%, and we concluded that the observed correlation was tighter than the expected one from the observational detection bias.

\begin{figure}[t]
  \centering
  \includegraphics[width=9cm]{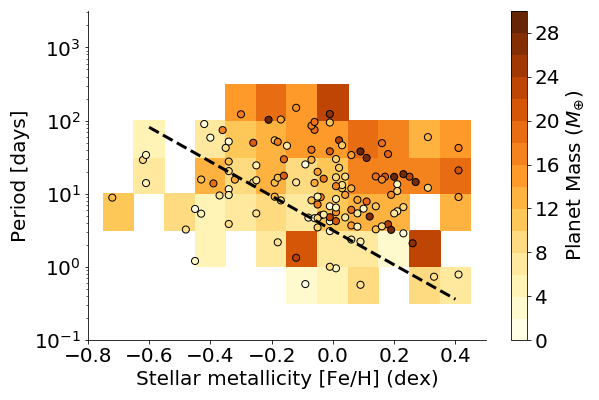} 
  \caption{Updated metallicity-period-mass diagram for low-mass planets ($ < 30 M_{\oplus}$) with planetary masses derived with at least 20\% precision. The color scheme represents the planet mass (in M$_{\oplus}$). The colored background corresponds to the diagram binned average planet mass. The dashed black line is the linear fit of the data for a planet mass of 10M$_{\oplus}$.}
  \label{fig_lmf_diag}
\end{figure}

Figure \ref{fig_lmf_diag} shows the metallicity-period-mass diagram for the updated SWEET-Cat sample (equivalent to Figure 2 in \citealt[][]{Sousa-2019}). With the new homogeneous parameters introduced in this work, we increase the low-mass planet sample with precise masses (within 20\%) to 131 planets (an increase of 56\%). We did a new 3D plane fit with the new low-mass planet sample, including the errors of all variables (see Appendix A of \citealt[][]{Sousa-2019}). We found the following metallicity-period-mass relation:


\begin{equation}
  M_{p} = (4.9 \pm 0.2) + \\
              (13.4 \pm 0.4) * [Fe/H] + \\
              ( 5.3 \pm 0.2) * \log(P)
\label{eq_mass}
.\end{equation}

This relation, although close to the one presented before, shows a slightly stronger dependence on both metallicity and the period. To assess the observational bias in this relation we compare these dependencies with the values extracted from the Monte Carlo simulations done in Section 3.2 in \citet[][]{Sousa-2019}; the expected observational bias correlation would produce a metallicity slope (m1) of 1.20 $\pm$ 8.43 [M$_{\oplus}$/dex] and a period slope (m2) of 3.21 $\pm$ 1.38 [M$_{\oplus}$/dex]. From these respective distributions the metallicity-mass correlation (m1) due to the observational bias is weak, but with high uncertainty. 

Our metallicity-mass correlation (13.4 [M$_{\oplus}$/dex]) is outside the 1$\sigma$ standard deviation, making it unlikely to retrieve this value by chance under our assumptions. For the period correlation (m2) the expected correlation due to observational bias is high, but more precise when compared with m1. The value for m2, 5.3 [M$_{\oplus}$/dex] obtained with the updated observational data is still close, but also now it appears to be outside  1$\sigma$  in comparison with the previous results. 

Compared with previous results, the increased number of low-mass exoplanets strengthens the linear correlations of planet mass with metallicity and period. However, we still cannot exclude the possibility that the observation bias can explain the correlations found.

\section{Summary}

We presented a significant update of SWEET-Cat where 283 planet host stars have new precise and homogeneous spectroscopic parameters using spectra that we collected from different sources. We reviewed nearly all spectroscopic parameters using new combined spectra and recent versions of the codes used in our spectroscopic analysis. We  also updated SWEET-Cat with relevant columns for the stellar characterization, most of them related to GAIA eDR3. We used the GAIA parallaxes and photometry to estimate accurate trigonometric surface gravities to include them in SWEET-Cat. We also reviewed planet--host metallicity correlations with a specific focus on the differences between stars hosting low-mass planets and stars hosting massive ones. The well-established correlation for stars hosting giant planets continues to hold, while with the increased number of low-mass planet detections, we were able to better characterize the planet--metallicity correlation for the low-mass regime. We found that stars hosting low-mass planets have a similar distribution to the stars in the solar neighborhood, but that slightly higher metallicities were favored. We showed the metallicity distributions for massive planets, splitting them between hot and cool according to their period. We found differences between these distributions that are consistent with results in the literature. Finally, we reviewed the metallicity-period-mass correlation for the low-mass planets. The correlation became slightly tighter with the inclusion of new planets, although it is still unclear if the reason behind it is the RV observational bias.

We will continue working on SWEET-Cat in the following years to provide the community with a data set relevant for statistical studies of exoplanets. There is still a significant number of relatively bright planet-hosts listed without homogeneous parameters. We will provide spectroscopic parameters once high-quality spectra are available. Furthermore, we have plans to include homogeneous element abundances for the stars with the highest quality spectra, make the spectra available for download on the website, and include more robust determinations of stellar masses, radii, and ages from stellar modeling.

\begin{acknowledgements}

This  work  was  supported  by FCT - Fundac\c{c}\~ao  para  a  Ci\^encia  e  Tecnologia through national  funds and by FEDER through COMPETE2020 - Programa Operacional Competitividade e Internacionaliza\c{c}\~ao by these grants: UID/FIS/04434/2019; UIDB/04434/2020; UIDP/04434/2020; PTDC/FIS-AST/32113/2017 \& POCI-01-0145-FEDER-032113; PTDC/FIS-AST/28953/2017 \& POCI-01-0145-FEDER-028953; PTDC/FIS-AST/28987/2017 \& POCI-01-0145-FEDER-028987; PTDC/FIS-OUT/29048/2017 \& IF/00852/2015. S.C.C.B. acknowledges support from FCT through contract nr. IF/01312/2014/CP1215/CT0004. S.G.S acknowledges the support from FCT through Investigador FCT contract nr. CEECIND/00826/2018 and  POPH/FSE (EC). V.A.   and EDM acknowledges  the  support  from  FCT  through  Investigador  FCT  contract nrs.  IF/00650/2015/CP1273/CT0001 and IF/00849/2015/CP1273/CT0003. B.R-A acknowledges funding support from FONDECYT through grant 11181295. This research has made use of the NASA Exoplanet Archive, which is operated by the California Institute of Technology, under contract with the National Aeronautics and Space Administration under the Exoplanet Exploration Program. This work has made use of data from the European Space Agency (ESA) mission {\it Gaia} (\url{https://www.cosmos.esa.int/gaia}), processed by the {\it Gaia} Data Processing and Analysis Consortium (DPAC, \url{https://www.cosmos.esa.int/web/gaia/dpac/consortium}). Funding for the DPAC has been provided by national institutions, in particular the institutions participating in the {\it Gaia} Multilateral Agreement. This work has been carried out within the framework of the NCCR PlanetS supported by the Swiss National Science Foundation. SH acknowledges CNES funding through the grant 837319.
\end{acknowledgements}

\bibliographystyle{aa}
\bibliography{sousa_bibliography}

\begin{appendix} 

\section{Spectroscopic parameters using the trigonometric surface gravity}

\begin{figure*}
  \centering
  \includegraphics[width=18cm]{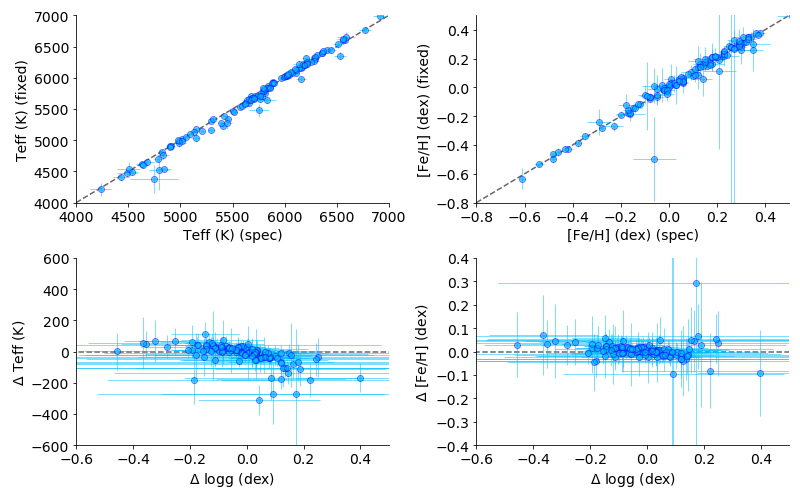}
  \caption{Temperature and metallicity derived with spectroscopic (spec) and trigonometric surface gravity (fixed). Top panels: Direct comparison of the spectroscopic effective temperatures and metallicities. Bottom panels: Differences in the effective temperature and metallicity vs. the difference in surface gravity (defined as trigonometric - spectroscopic).}
  \label{logg_fixed_deps}
\end{figure*}


Figure \ref{logg_fixed_deps} shows the effective temperatures and metallicities derived by the standard analysis against the same parameters using trigonometric surface gravities. We used a subsample of 150 planet-host stars randomly selected from the whole homogeneous sample in SWEET-Cat. From these stars, 117 hosts   converged results in a single run of the spectroscopic analysis ARES+MOOG when constraining the surface gravity. These constrained parameters are the ones used in Figure \ref{logg_fixed_deps}. As expected, the effective temperatures and metallicities show small changes using the constrained surface gravity. The mean differences between the effective temperatures and the metallicity values are -20 $\pm$ 76 K and 0 $\pm$ 0.06 dex, respectively. These values were derived even with the presence of a few outliers noticeable in the figure. These are stars with a low S/N spectrum or stars cold enough to make difficult the derivation of reliable parameters \citep[e.g.,][]{Tsantaki-2013}. In the bottom panel of Figure \ref{logg_fixed_deps} is shown the dependence of the differences on the changes in the surface gravity. There are evident trends in these plots, but they show little dependence on the differences in the surface gravity derived in the ARES+MOOG. The above is compatible with Figures 4 and 5 of \citet[][]{Mortier-2014}, even when that sample includes evolved stars. Therefore, the unconstrained spectroscopic analysis ARES+MOOG derives reliable and consistent effective temperatures and metallicities, even in the cases where we have less accurate spectroscopic surface gravities.

\section{Table with SWEET-Cat contents}

\begin{table*}
\caption[]{New contents of SWEET-Cat}
  \begin{center}

    \begin{tabular}{lll}
    \hline
    \hline
    Column     &   Units  &    Description  \\
    \hline

Name         &  -       &  Identifier name for the planet-host star \\
hd           &  -       &  HD Number \\
RA           & [h:m:s]  &  Right Ascension (ep=J2000)\\ 
DEC          & [deg:m:s]&  Declination (ep=J2000)\\ 
Vmag         &  -       &  Visual magnitude \\ 
eVmag        &  -       &  Error on visual magnitude \\ 
PlxFlag      &  -       &  Flag indicating source of parallax \\ 
Teff         &  [K]     &  Effective temperature \\ 
eTeff        &  [K]     &  Error on effective temperature. \\
Logg         &  [dex]   &  Spectroscopic surface gravity. \\ 
eLogg        &  [dex]   &  Error on spectroscopic surface gravity. \\ 
Vt           &  [km/s]  &  Microturbulence. \\ 
eVt          &  [km/s]  &  Error on microturbulence. \\

[Fe/H]       &  [dex]   &  Iron abundance (proxy for metallicity). \\ 
e[Fe/H]      &  [dex]   &  Error on iron abundance. \\ 
Reference    &  -       &  Source of the spectroscopic parameters. \\ 
Link         &  -       &  URL address to reference. \\       
SWFlag       &  -       &  Homogeneous flag. \\ 
Update       &  -       &  Date of last update on entry. \\ 
Comment      &  -       &  Relevant comments for entry. \\ 
Database     &  -       &  Indicates if planet host is in exo.eu and/or NASA archive \\ 
gaia\_dr2    &  -       &  GAIA DR2 source id. \\ 
gaia\_dr3    &  -       &  GAIA eDR3 source id. \\
Plx          &  [mas]   &  Parallax \\ 
ePlx         &  [mas]   &  Error on parallax \\ 
Gmag         &  -       &  GAIA G magnitude \\ 
eGmag        &  -       &  Error on GAIA G magnitude. \\ 
RPmag        &  -       &  GAIA red pass magnitude \\ 
eRPmag       &  -       &  Error on red pass magnitude. \\ 
BPmag        &  -       &  GAIA blue pass magnitude \\
eBPmag       &  -       &  Error on blue pass magnitude \\ 
FG           &  [e-/s]  &  Mean flux of GAIA G band \\ 
eFG          &  [e-/s]  &  Error on mean flux of GAIA G band \\ 
G\_flux\_std\_n  &-     &  GAIA photometry activity index \\ 
Logg\_gaia   &  [dex]   &  Derived GAIA eDR3 surface gravity \\ 
eLogg\_gaia  &  [dex]   &  error on derived GAIA eDR3 surface gravity. \\Mass\_t      &  [M$_\odot$]  &  Derived stellar mass. \\
eMass\_t     &  [M$_\odot$]  &  Error on derived stellar mass. \\
Radius\_t    &  [R$_\odot$]  &  Derived stellar radius. \\
eRadius\_t   &  [R$_\odot$]  &  Error on derived stellar radius. \\
spec\_base   &  -            &  Base name of spectra filename. \\
Distance     &  [pc]         &  Distance - inverted from Parallax \\
Distance\_b  &  [pc]         &  Distance from \citet[][]{Bailer-Jones-2021} \\
eDistance\_b &  [pc]         &  Error on distance from \citet[][]{Bailer-Jones-2021} \\
RA\_EU       &  [deg]        &  Right Ascension (exo.eu) \\ 
DEC\_EU      &  [deg]        &  Declination (exo.eu) \\ 
RA\_NASA     &  [deg]        &  Right Ascension (NASA archive) \\ 
DEC\_NASA    &  [deg]        &  Declination (NASA archive) \\

\hline
    \end{tabular}
  \end{center}
\label{tab_contents}
\end{table*}

\end{appendix}

\end{document}